\documentclass[12pt,a4paper,twoside]{article}
\usepackage{epsfig}

\newcommand{\be}{\begin{equation}}
\newcommand{\ee}{\end{equation}}
\newcommand{\beqs}{\begin{eqnarray}}
\newcommand{\eeqs}{\end{eqnarray}}

\newcommand{\tr}{{\rm tr}}
\newcommand{\Tr}{{\rm Tr}}
\newcommand{\half}{{1 \over 2}}
\newcommand{\ba}{{\bf a}}
\newcommand{\bb}{{\bf b}}
\newcommand{\tha}{{\cal \theta}_1}
\newcommand{\thh}{{\cal \theta}_4}

\def\NP{{\it Nucl. Phys.\ }}
\def\PL{{\it Phys. Lett.\ }}

\def\PR{{\it Phys. Rev.\ }}
\def\PRL{{\it Phys. Rev. Lett.\ }}

\def\JMP{{\it J. Math. Phys.\ }}

\def\MPL{{\it Mod. Phys. Lett. A\ }}

\def\LNC{{\it Lett. Nuovo Cimento \ }}

\expandafter\ifx\csname mathbbm\endcsname\relax

\else

\fi
\begin{document}
\begin{titlepage}
\begin{flushleft}  
       \hfill                      {\tt hep-th/9810211}\\
       \hfill                      UUITP-8/98\\
       \hfill                       October 1998\\
\end{flushleft}
\vspace*{3mm}
\begin{center}
{\LARGE Generalized Calogero models through reductions
by discrete symmetries \\}
\vspace*{12mm}
\large Alexios P. Polychronakos\footnote{E-mail:
poly@teorfys.uu.se} \\
\vspace*{5mm}
{\em Institutionen f\"{o}r teoretisk fysik, Box 803 \\
S-751 08  Uppsala, Sweden \/}\\
\vspace*{4mm}
and\\
\vspace*{4mm}
{\em Physics Department, University of Ioannina \\
45110 Ioannina, Greece\/}\\
\vspace*{15mm}
\end{center}

\begin{abstract}
We construct generalizations of the Calogero-Sutherland-Moser
system by appropriately reducing a classical Calogero model by
a subset of its discrete symmetries. Such reductions reproduce 
all known variants of these systems, including some recently
obtained generalizations of the spin-Sutherland model,
and lead to further generalizations of the elliptic model
involving spins with $SU(n)$ non-invariant couplings.

\end{abstract}

\vspace*{10mm}

\end{titlepage}

The inverse-square interacting particle system \cite{Cal,Suth,Mos} 
and its spin generalizations \cite{GH,Woj,HH,Kaw,MP1,HW} are important 
models of many-body systems, due to their exact solvability and
intimate connection to spin chain systems \cite{Hal,Sha,FM,Scal},
2-dimensional Yang-Mills theories \cite{GN,MP2,LSK} etc.

Most of the variants of these systems can be though of as appropriate
`foldings' of the basic Calogero model with an augmented number of
particles. Versions of this idea have appeared in the
early literature, and have been been used, e.g., to motivate the
Sutherland \cite{Suth} and elliptic (Weierstrass) \cite{OP}
versions of these systems.
In this paper, we use this approach in the case of spin-generalized
systems to give a more intuitive derivation and interpretation to some
recently produced models and to derive new models.

We begin with a brief review of known cases. We consider the
Calogero model in connection with some of its discrete symmetries $D$.
The equations of motion remain invariant under the phase space mapping 
$\phi \to D(\phi)$, where $\phi$ are phase space variables. Then the
reduction to the invariant subspace $\phi = D(\phi)$ is kinematically
preserved, that is, the equations of motion do not move the system out
of this subspace. Therefore, reducing the initial value data to this
subspace trivially produces a system as solvable as the original one.
The motion will be generated by the original hamiltonian on the 
reduced space.

The starting point will be the inverse-square scattering particle system
\be
H = \sum_{i=1}^N \half p_i^2 + \half \sum_{i \neq j} 
{g \over x_{ij}^2}
\ee
where $x_i$, $p_i$ are particle coordinates and momenta in one dimension
and $x_{ij} = x_i - x_j$.  The considered symmetries are:

$\bullet$ Translation invariance $T$: $x_i \to x_i + \ba$, $p_i \to p_i$

$\bullet$ Parity $P$: $x_i \to - x_i$, $p_i \to - p_i$

$\bullet$ Permutation symmetry $M$: $x_i \to x_{M(i)}$, $p_i \to p_{M(i)}$
with $M$ any element of the permutation group of $N$ particles.

Other symmetries will not be useful for our purposes.

A direct reduction of the system by any of the above symmetries
does not produce anything nontrivial or sensible: $\phi = T(\phi)$
is possible only in the trivial case $a=0$, while $\phi = P(\phi)$
and $\phi = M(\phi)$ requires (some) of the particle coordinates to
coincide, which is excluded by the infinite two-body potential.
We get useful systems only when reducing through appropriate products
of the above symmetries. These are:

\vskip 0.2cm

a) $D=PM$: We reduce by $P$ and a particular permutation: $M(i) = N-i+1$
(or any other in the same conjugacy class). This is uniquely fixed
by the requirement that $M^2 =1$ (so that $D^2 (\phi) = M^2 (\phi)$
does not make any two different particle coordinates to coincide)
and that $M(i) = i$ for at most one $i$ (so that no two or
more particle coordinates are put to zero). The constraint: 
\be
x_i = -x_{N-i+1} ~,~~~p_i = -p_{N-i+1}
\ee
effectively reduces the 
original system into two mirror-images. The reduced hamiltonian is
\begin{eqnarray}
H &=& \sum_{i=1}^{N'} \half p_i^2 
+ \half \sum_{i \neq j} {g \over (x_i - x_j )^2}
+ \half \sum_{i \neq j} {g \over (x_i + x_j )^2}
+ \sum_i {g' \over x_i^2} \nonumber \\
N' &=& \left[\frac{N}{2} \right] ~,~~~ g' = g \left(\frac{1}{4} 
+ 2\left\{ \frac{N}{2} \right\} \right)
\end{eqnarray}
where an overall factor of $2$ has been discarded and $[.]$, $\{ . \}$
denote integer and fractional part, respectively.
The second term in the potential is the interaction
of each particle with the mirror image of each other particle; the
third part accounts for the interaction of each particle with the
mirror image of itself, and with a particle fixed at the origin by the
constraint (for odd $N$).

We further note that parity symmetry persists in the case where 
an external harmonic oscillator potential is added to the system,
promoting it to the confining, rather than scattering, Calogero model:
\be
H = \sum_{i=1}^N \half p_i^2 + \half \sum_{i \neq j} 
{g \over x_{ij}^2} + \sum_{i=1}^N \half \omega^2 x_i^2
\ee
Reduction by $D=PM$ produces an integrable system similar as above with
the added harmonic oscillator potential.

\vskip 0.2cm

b) $D=TM$: No finite-rank element of the permutation group will do, 
since repeated application of $D$ would eventually lead to $x_i = x_i +m\ba$.
We overcome this by starting with $N'$ particles and taking the limit 
$N' \to \infty$. We pick the element: $M(i) = i+N$ for some finite $N$ 
which, for infinite $N'$, is infinite-rank. The constraint: 
\be
x_{i+N} = x_i +\ba ~,~~~ p_{i+N} = p_i
\ee
leads to a system consisting of infinitely many
copies of a finite system displaced by multiples of $a$. 
We can parametrize the particle indices by the pair $(i,m)$,
$i =1 \dots N$ and $m \in {\bf Z}$, where the original index
is $i+mN$. The constraint now reads
\be
x_{i,m} = x_i +m\ba ~,~~~ p_{i,m} = p_i
\label{conb}
\ee
The resulting system is infinite copies of an $N$-body system.
The reduced hamiltonian is
\begin{eqnarray}
H &=& \sum_{m=-\infty}^\infty \sum_{i=1}^N  \half p_i^2 + 
\half \sum_{m,n=-\infty}^\infty \sum_{i,j} 
{g \over (x_{ij} +m\ba -n\ba)^2}
\nonumber \\
&=& \half \sum_{m'=-\infty}^\infty \left\{\sum_{i=1}^N p_i^2 + 
\sum_{m=-\infty}^\infty \sum_{i \neq j} 
{g \over (x_{ij} +ma )^2} + \sum_i \sum_{m \neq 0}
{g \over (ma)^2} \right\}
\end{eqnarray}
In the above summation terms with ($i=j$, $m=n$) are omitted since
they correspond to self-interactions
of particles that are excluded from the original Calogero model.
The summation over $m'$ above accounts for the infinite periodically
repeating copies of the system and can be dropped. 
The infinite $m$-summation accounts for the interaction of each 
particle with the multiple images of each other particle and can
be performed explicitly. We eventually get
\be
H = \sum_{i=1}^N \half p_i^2 + \half 
\sum_{i \neq j} {g \pi^2 \over \ba^2 \sin^2 \pi {x_{ij} \over \ba}}
\ee
that is, the Sutherland model. 
In the above we omitted an irrelevant constant term equal to 
$gN {\pi^2 \over 6\ba^2}$ coming from terms with $i=j$, which account
for the interaction of each particle with its own infinite images.

\vskip 0.2cm

c) We can formally extend the $T$ symmetry to complex parameter $\ba$.
As long as there is a subset of coordinates in the reduced phase
space that remains real and generates all other coordinates through
use of $D$, we will have a well-defined real subsystem. Applying
$D=TM$ for infinitely many particles parametrized by a double index
$(i,j)$, $i,j \in {\bf Z}$, for two complex translations $a$ and $b$,
the constraint is 
\be
x_{i,j} = x_{i+N',j} +\ba = x_{i,j+M'} +\bb ~,~~~
p_{i,j} = p_{i+N',j} = p_{i,j+M'} 
\ee
and we end up with a finite system with $N = M'N'$ particles periodically
repeating on the complex plane. Similarly to the Sutherland case,
the hamiltonian becomes infinitely many copies of
\be
H = \sum_{i=1}^{N'} \half p_i^2 + \half \sum_{i,j} 
\sum_{m,n=-\infty}^\infty {g \over (x_{ij} +m\ba+n\bb)^2}
\ee
The above sum has a logarithmic ambiguity that is easily regulated
by subtracting the constant $g/(m\ba+n\bb)^2$ from each term.
We end up with
\be
H = \sum_{i=1}^{N'} \half p_i^2 
+ \half \sum_{i \neq j} g {\cal P}(x_{ij} |\ba,\bb) 
\ee
that is, the elliptic model with Weierstrass potential.

\vskip 0.2cm

d) $D_1=PM$ together with $D_2=TM$: 
This is a combination of (a) and (b) above. We work
again with an infinite number of particles. We impose two constraints:
\begin{eqnarray}
x_{-j+1-\epsilon} = -x_j ~&,&~~~ p_{-j+1-\epsilon} = -p_j
\nonumber \\
x_{j+N'} = x_j +\ba ~&,&~~~ p_{j+N'} = p_j
\end{eqnarray}
where $\epsilon = 0,1$ (any other choice of $\epsilon$ is equivalent
to one of these).
Parametrizing $j=i+mN'$ by the pair $i,m$, $i=1 \dots N'$, $m \in 
{\bf Z}$, we have
\begin{eqnarray}
x_{N-i+1-\epsilon} = \ba-x_{i,m} ~&,&~~~ p_{N-i+1-\epsilon} = -p_i
\nonumber \\
x_{i,m} = x_i +m\ba ~&,&~~~ p_{i,m} = p_i
\end{eqnarray}
so we end up with a finite system of $N=[(N' -\epsilon)/2]$ particles. The 
reduced hamiltonian is infinitely many copies of
\begin{eqnarray}
H = \sum_{i=1}^{N} \half p_i^2 &+& \half 
\sum_{i \neq j} {g \pi^2 \over \ba^2 \sin^2 \pi {x_i - x_j \over \ba}}
+ \half 
\sum_{i \neq j} {g \pi^2 \over \ba^2 \sin^2 \pi {x_i + x_j \over \ba}}
\nonumber \\
&+& \half 
\sum_{i} {g' \pi^2 \over \ba^2 \sin^2 \pi {2x_i \over \ba}}
+ \half 
\sum_{i} {g'' \pi^2 \over \ba^2 \sin^2 \pi {x_i \over \ba}}
\end{eqnarray}
with
\be
g' = g \left( \frac{1}{2} + 8\left\{ \frac{N-\epsilon}{2} \right\}
\right) ~,~~~
g'' = g \left( \epsilon - 2\left\{ \frac{N-\epsilon}{2} \right\}
\right)
\label{gg}
\ee
The coordinates $x_i$ can all be taken in the interval $(0,\ba/2)$
and the particles interact with their infinite mirror-images with
respect to mirrors placed at $x=0$ and $x=\ba/2$ and with
particles fixed at $x=0$ (if $\epsilon=1$) and at $x=\ba/2$
(if $2\{ (N-\epsilon)/2 \}=1$).
A similar construction can be performed with two complex translations,
as in (c) plus one parity reversal. We obtain a similar model
but with elliptic functions appearing instead of inverse sine squares.

\vskip 0.4cm

The above is fairly standard and exhausts the possibilities for
spinless particles. Before we proceed to the more novel and
interesting case of particles with spin, we find it instructive to
demonstrate how the above construction reproduces the 
conserved integrals of motion of the reduced system. We will consider 
case (b), as the most generic, case (a) being rather trivial.

The Lax matrix of the original scattering Calogero model is \cite{OP}
\be
L_{ij} = p_i \delta_{ij} + (1-\delta_{ij} ) \frac{i\ell}
{x_{ij}}
\ee
where $\ell^2 = g$. Traces of powers of $L$ produce the integrals
of motion in involution for the model:
\be
I_k = \tr L^k ~,~~~k=1,\dots N
\ee

For the system of case (b), we promote the index $i$ into a
pair $(i,m)$ and choose $x_{i,m} = x_i + m\ba$ as in (\ref{conb}).
The resulting infinite-dimensional matrix $L_{im,jn}$ can be thought of
as consisting of infinitely many blocks of size $N \times N$,
$m,n$ labeling the blocks and $i,j$ the elements of each block:
\be
L_{im,jn} = p_i \delta_{mn} \delta_{ij}
+ (1-\delta_{mn} \delta_{ij} ) \frac{i\ell}{x_{ij} + (m-n)\ba}
\ee
We observe a block `translational invariance' of the matrix $L$ in
the indices $m,n$, which reflects the invariance of the model under 
a translation by $\ba$. Due to this, we can trade the pair
$m,n$ for a single index $m-n$ 
\be
L_{im,jn} \equiv L_{m-n;ij}
\ee
and thus $L$ becomes an infinite
collection of $N \times N$ matrices $L_n$ labeled by $n$.
We define the Fourier transform $L(\sigma )$:
\be
L(\sigma) = \sum_n e^{in\sigma} L_n 
\ee
in terms of which $L_n$ is
\be
L_n = \frac{1}{2\pi} \int_0^{2\pi} L(\sigma) e^{-in\sigma}
\ee
The corresponding integrals of motion $I_k$ are traces of powers
of $L$. Denoting by $\Tr$ the trace in the infinite-dimensional
space labeled by $i,m$ and by $\tr$ the trace in the $N$-dimensional
space labeled by $i$ alone, we have:
\begin{eqnarray}
I_k = \Tr L^k &=& \sum_{n_1 , \dots n_k} \tr( L_{n_1 -n_2} 
\cdots L_{n_k -n_1} ) \nonumber \\
(m_i \equiv n_i - n_{i+1})~~~
&=& \sum_{n_1} \, \sum_{m_1 , \dots m_{k-1}}
\tr ( L_{m_1} \cdots L_{m_{k-1}} L_{-m_1 \cdots -m_{k-1}} )
\end{eqnarray}
The sum over $n_1$ above produces a trivial infinity. This is
due to the summation over the infinite copies of the system,
just as in the case of the hamiltonian, and will be dropped.
In terms of the Fourier transformed $L(\sigma)$ the reduced
$I_k$ become simply
\be
I_k = \frac{1}{2\pi} \int_0^{2\pi} \tr L(\sigma)^k d\sigma
\label{Ik}
\ee
It is now a matter of calculating $L(\sigma)$. From the
Fourier transform
\be
\sum_n \frac{e^{in\sigma}}{n+x} = \frac{\pi e^{i(\pi -\sigma)x}}
{\sin \pi a} ~~~{\rm for}~~0<\sigma<2\pi
\label{sum}
\ee
we obtain for $L(\sigma)$
\be
L(\sigma)_{ij} = e^{i(\pi-\sigma)x_{ij}} \left[
p_i \delta_{ij} + (1-\delta_{ij})\frac{i\pi\ell}{\ba \sin\pi
\frac{x_{ij}}{\ba}} -\ell\frac{\pi-\sigma}{\ba}\delta_{ij} \right]
\label{Ls}
\ee
where the last diagonal term linear in $\pi-\sigma$ came from
terms with $i=j$, $n \neq 0$ in $L_{n,ij}$. We observe that the
matrix inside the square bracket apart from this linear part
is the standard Lax matrix $\tilde L$ of the Sutherland model:
\be
{\tilde L}_{ij} = p_i \delta_{ij} + (1-\delta_{ij})
\frac{i\pi\ell}{\ba \sin\pi\frac{x_{ij}}{\ba}} 
\ee

Substituting (\ref{Ls}) in (\ref{Ik}) we note that the
exponential factors cancel (due to $x_{i_1 i_2} + \cdots x_{i_{k-1}
i_1} =0$) and we are left with
\begin{eqnarray}
I_k &=& \frac{1}{2\pi} \int_0^{2\pi} \tr \left( {\tilde L} -\ell \frac{
\pi-\sigma}{\ba} \right)^k d\sigma
= \sum_{s=0}^k \tr \frac{k!}{s! (k-s)!} {\tilde L}^{k-s} 
\frac{(-\ell)^s}{2\pi \ba^s} 
\int_0^{2\pi} (\pi-\sigma)^s d\sigma \nonumber \\
&=& \sum_{n=0}^{[k/2]} 
\frac{k!}{(2n+1)! (k-2n)!} 
\left( \frac{\pi \ell}{\ba} \right)^{2n}
~ {\tilde I}_{k-2n} 
\end{eqnarray}
where ${\tilde I}_k = \tr {\tilde L}^k$ are the conserved integrals
of the Sutherland model. We obtain a linear combination
of the intergal ${\tilde I}_k$ and lower integrals of the same parity.
The appearence of the lower integrals originates from the interaction
of each particle with its own infinite images. We saw an example of
such a term in the constant potential term that we omitted from the
reduced hamiltonian of case (b). In conclusion, we have 
recovered the integrals of the Sutherland model.

\vskip 0.4cm

We extend now these considerations to systems of particles with internal 
classical $U(n)$ degrees of freedom. The 
corresponding starting spin-Calogero system can be obtained,
for instance, from the model in \cite{GH,Woj} (which can
itself be obtained as a reduction of a hermitial matrix model
\cite{KKS} into nontrivial angular momentum sectors) 
by redistributing the global $U(M)$ degrees of freedom of this model
into individual particle `spins'. Equivalently, we can take the 
infinite-volume classical limit of the spin model derived and solved
in \cite{MP2}. The hamiltonian reads
\be
H = \sum_{i=1}^N \half p_i^2 + \half \sum_{i \neq j} {\tr(S_i S_j )
\over x_{ij}^2}
\ee
The $S_i$ are a set of independent classical $U(n)$ spins of rank one 
and length
$\ell$, that is, $n \times n$ rank-one hermitian matrices satisfying
\be
\tr (S_i )^2 =  \ell^2
\label{lengthl}
\ee
and with Poisson brackets
\be 
\{ (S_i)_{ab} , (S_j)_{cd} \} = -i \delta_{ij}
\left[ (S_i)_{ad} \delta_{cb} - \delta_{ad} (S_i)_{cb} \right]
\ee
Such spins can be realized in terms of oscillators \cite{MP2}: 
\be
(S_i )_{ab} = {\bar A}_i^a A_i^b
~,~~ a,b=1 \dots n
\label{Sa}
\ee
where $( A_i^a \,,\, {\bar A}_i^a)$ are a set of $nN$ independent
classical harmonic oscillator canonical pairs with Poisson brackets:
\be
\{ A_i^a , {\bar A}_j^b \} = i \delta_{ij} \, \delta_{ab}
\ee
and satisfying the constraint
\be
\sum_a {\bar A}_i^a A_i^a = \ell ~~{\rm for~all}~i
\label{Acon}
\ee

The above model, in addition to the previous symmetries $T$, $P$ and
$M$, also possesses the symmetry

$\bullet$ Spin rotations $U$: $S_i \to U S_i U^{-1}$, with $U$ a 
constant unitary $n \times n$ matrix. $(x_i , p_i )$ remain unchanged.

Again, reduction by this symmetry alone leads to no interesting
system (implying either $U=1$ or $S_i=0$). Reduction by $PUM$
or $TUM$, however, much along the lines of the previous $PM$ 
and $TM$ reductions, produces new and nontrivial results:

\vskip 0.2cm

e) $D=PUM$ with $P$ and $M$ as in (a) before, and $U$ a unitary
matrix satisfying $U^2 =1$ (this is necessary since $P$ and $M$
are of rank two). The constraints are
\be
x_i = -x_{N-i+1} ~,~~~p_i = -p_{N-i+1} ~,~~~ S_i = U S_{N+i-1} U^{-1}
\ee
The reduced hamiltonian acquires the form:
\begin{eqnarray}
H = \sum_{i=1}^{N'} \half p_i^2 
&+& \half \sum_{i \neq j} {\tr(S_i S_j) \over (x_i - x_j )^2}
+ \half \sum_{i \neq j} {\tr(S_i U S_j U^{-1}) \over (x_i + x_j )^2}
\nonumber \\
&+& \sum_i {\tr(S_i U S_i U^{-1}) \over 4x_i^2}
+ g\sum_i {\tr(S_i S_o) \over x_i^2}
\end{eqnarray}
where $N' = [N/2]$, $g=2\{N/2\}$ and $S_o = U S_o U^{-1}$ 
is an extra spin degree of freedom.

The form of the hamiltonian for the reduced model and its physical
interpretation simplifies with an appropriate choice of basis for
the spins: by using the $U$-invariance of the full model, we can
perform a unitary rotation $V$ to all spins $S_i \to V S_i V^{-1}$.
This transforms the matrix $U$ appearing in (\ref{TUMd}) into
$U \to V^{-1} U V$. With an appropriate choice of $V$ we can always
choose $U$ to be diagonal: $U = diag (e^{i \phi_a})$. Because of
the constraint, this means that $U$ will have the form
$U = diag(1,\dots 1,-1,\dots -1)$ with $n_1$, $n_2$ entries equal
to $1$, resp. $-1$. So we see that the original $U(n)$ invariance
of the model has been broken to $U(n_1) \times U(n_2)$. If $n_1
= n_2$ there is an additional $Z_2$ exchange symmetry.

As in the spinless case (a), we could have started with a spin-Calogero
model in an external oscillator potential (which shares the same
$U$ and $P$ symmetries), and obtain a model as above with the
extra confining harmonic potential.

\vskip 0.2cm

f) $D=TUM$ with $T$ and $M$ as in (b) before, and $U$ any unitary
matrix. The constraint on the phase space is
\be
x_i = x_{i+N} +\ba ~,~~~ p_i = p_{i+N} ~,~~~ S_{i+N} = U S_i U^{-1}
\label{TUMd}
\ee
The system becomes, again, infinite copies of $a$-translated and
$U$-rotated systems, and the reduced hamiltonian is
\be
H = \sum_{i=1}^N \half p_i^2 + \half \sum_{i,j} 
\sum_{m=-\infty}^\infty {\tr(S_i U^m S_j U^{-m} ) 
\over (x_{ij} +m\ba )^2}
\label{HTUMd}
\ee
In the above we cannot drop terms with $i=j$ any more, since they are
now spin-dependent rather than constant. Only the term ($i=j$, $m=0$)
must be dropped from the summation as before.

Again, the form of the hamiltonian for the reduced model and its physical
interpretation simplifies with an appropriate choice of basis for
the spins which makes $U$ diagonal: $U = diag (e^{i \phi_a})$. 
The trace in (\ref{HTUMd}) then becomes
\be
\tr ( S_i U^m S_j U^{-m} ) = \sum_{a,b=1}^n (S_i)_{ab} (S_j)_{ba}
e^{-im \phi_{ab}}
\ee
where $\phi_{ab} = \phi_a - \phi_b$. The 
$m$-summation appearing in (\ref{HTUMd}) gives
\begin{eqnarray}
\sum_{m=-\infty}^\infty {\tr(S_i U^m S_j U^{-m} )\over 
(x_{ij} +m\ba )^2} &=&
\sum_{a,b=1}^n (S_i)_{ab} (S_j)_{ba} \sum_{m=-\infty}^\infty
{e^{-im \phi_{ab}} \over (x_{ij} +m\ba )^2 } \\
&=& \sum_{a,b=1}^n V_{ab} (x_{ij}) \, (S_i)_{ab} (S_j)_{ba} 
\end{eqnarray}
with the potential $V_{ab} (x)$ being
\be
V_{ab} (x) = \sum_{m=-\infty}^\infty
{e^{-im \phi_{ab}} \over (x +m\ba )^2 } 
\ee
We must distinguish between the cases $i \neq j$ and $i=j$.
For the case $i \neq j$ the sum can be obtained from the
$x$-derivative of (\ref{sum}):
\be
V_{ab} (x) = \frac{1}{\ba^2} e^{-i{x \over \ba}\phi_{ab}}
\left( {\pi^2 \over \sin^2{\pi x \over \ba}} -i\pi \phi_{ab}
\cot{\pi x \over \ba} - \pi | \phi_{ab} | \right)
\ee
For the case $i=j$ we must omit the term $m=0$ from the summation.
We obtain an $x$-independent potential:
\be
{\tilde V}_{ab} \equiv \lim_{x \to 0} \left( V_{ab} (x) -
\frac{1}{x^2} \right) = {\phi_{ab}^2 \over 2\ba^2} - 
{\pi | \phi_{ab} | \over \ba^2}
\ee
In the above we omitted a constant ($a,b$-independent) term equal
to $\frac{\pi^2}{3\ba^2}$ which would contribute to the hamiltonian
a term proportional to $\sum_i \tr (S_i )^2$. Due to (\ref{lengthl}), 
this is an irrelevant constant. 
With the above, the reduced hamiltonian eventually becomes 
\be
H = \sum_{i=1}^N \half p_i^2 + \half \sum_{i \neq j} 
\sum_{a,b} V_{ab} (x_{ij}) \, (S_i)_{ab} (S_j)_{ba} +
\half \sum_i \sum_{a,b} {\tilde V}_{ab} \, (S_i)_{ab} (S_i)_{ba} 
\label{Hab}
\ee
This is a model of particles with $U(n)$ spins interacting through
$U(n)$ {\it non}-invariant couplings, due to the presence of the
matrix $V_{ab}$. The original global $U(n)$ invariance is, now, broken
to the diagonal $U(1)^n$ part and only the diagonal components $S_{aa}$
of the total spin
\be
S_{ab} = \sum_i (S_i)_{ab}
\ee
are conserved.
The standard $U(n)$-invariant spin-Sutherland model is recovered
upon choosing $V_{ab} \sim \delta_{ab}$, in which case the sums
over $a,b$ above become a normal trace. This is achieved by choosing
$\phi_a$=constant, that is, $U=e^{i\phi}$.

The above model is, in fact, the same as the classical model
introduced by Blom and Langmann \cite{BL}, and this author
\cite{AP}, in the particle-spin form in which it was recast
in \cite{AP}:
\begin{eqnarray}
H &=& \half \sum_i p_i^2 
+ \half \sum_{i \neq j} \left( \sum_{ij} V_{ab} (x_{ij} ) 
({\hat S}_i)_{ab} ({\hat S}_j)_{ba} + {\ell (\ell +n) \over 4n 
\sin^2 {x_{ij} \over 2}}\right) \\
&+& \half \sum_i \sum_{ab} {\tilde V}_{ab} \, ({\hat S}_i)_{aa} 
({\hat S}_i)_{bb} + {1 \over 2N} \sum_{ab} {\tilde V}_{ab} \left( q_a q_b
- {\hat S}_{aa} {\hat S}_{bb} \right)
\label{Hsp}
\end{eqnarray}
To fully see the equivalence, we must observe
the following:

1. In the present construction we expressed the hamiltonian in terms
of $U(n)$ spins $S_i$. In \cite{AP} it was, instead, expressed in terms of
traceless $SU(n)$ spins ${\hat S}_i$. By (\ref{Sa}) and (\ref{Acon}) we 
have $\tr S_i = \ell$, so the relation between the two is
\be
{\hat S}_i = S_i - \frac{\ell}{n}
\ee

2. The expression (\ref{Hsp}) derived in \cite{AP} was fully
quantum mechanical. It can be seen that the term $\ell (\ell +n)$
in (\ref{Hsp}) classically becomes $\ell^2$ ($n$ was a quantum
correction similar to the shift of the classical angular momentum
$J^2$ to $J(J+1)$).

3. For the rank-one matrices $S_i$ we have the relation
\be
(S_i)_{ab} (S_i)_{ba} = (S_i)_{aa} (S_i)_{bb}
\ee

4. In \cite{AP} a set of dynamically conserved charges $q_a$
were introduced that can be chosen to have any value as long
as they sum to zero.

5. In \cite{AP} the particles were taken to move on the unit
circle, that is, $\ba=2\pi$.

Doing the above substitutions in (\ref{Hab}) we see that it becomes
practically identical to (\ref{Hsp}). The two expressions  differ
by constant terms depending on the charges $q_a$
and the diagonal elements of the total spin $S_{aa}$. Since both
of these quantities are constants of the motion, the two models
are trivially related.

\vskip 0.2cm

g) We can, similarly to (c), extend the above construction to two
complex translations and corresponding spin rotations. Parametrizing
again the infinite number of particles with a doublet of indices
$i,j \in {\bf Z}$ the constraints are
\begin{eqnarray}
x_{i,j} = x_{i+N,j} +\ba ~,~~~ p_{i,j} &=& p_{i+N,j} 
~,~~~ S_{i+N,j} = U S_{ij} U^{-i} \nonumber \\
x_{i,j} = x_{i,j+M} +\bb ~,~~~ p_{i,j} &=& p_{i,j+M} 
~,~~~ S_{i,j+M} = V S_{i,j} V^{-1}
\end{eqnarray}
and we end up as before with a finite system with $N' = MN$ 
particles periodically repeating on the complex plane. Since the
two space translations and the corresponding particle permutations
commute, for consistency the two spin rotations must also commute:
\be
UV = VU
\ee
The corresponding reduced hamiltonian becomes infinite copies
of 
\be
H = \sum_{i=1}^{N'} \half p_i^2 + \half \sum_{i,j} 
\sum_{m,n=-\infty}^\infty {\tr( U^m V^n S_i U^{-m} V^{-n} S_j ) 
\over (x_{ij} +m\ba+n\bb)^2}
\label{Hmn}
\ee
Just as in the previous case, we can choose a basis for the spins
that diagonalizes both $U$ and $V$ to $U=diag( \phi_i )$,
$V=diag( \theta_i )$. The $m,n$-sums that appear in (\ref{Hmn})
become
\be
\sum_{a,b=1}^n (S_i)_{ab} (S_j)_{ba} 
\sum_{m,n=-\infty}^\infty { e^{-im\phi_{ab} -in\theta_{ab}}
\over (x_{ij} +m\ba+n\bb)^2}
\ee
where the term $m=n=0$ is omitted if $i=j$. 
We obtain again
a potential $V_{ab} ( x_{ij} )$, for $i \neq j$, given by the sum
\be
V_{ab} (x) =
\sum_{m,n=-\infty}^\infty { e^{-im\phi_{ab} -in\theta_{ab}}
\over (x +m\ba+n\bb)^2}
\ee
and a spin self-coupling ${\tilde V}_{ab}$ for $i=j$, given by
\be
{\tilde V}_{ab} = 
\sum_{(m,n) \neq (0,0)} { e^{-im\phi_{ab} -in\theta_{ab}}
\over (m\ba+n\bb)^2}
\ee
Note that now, due to the presence of the phase factors, these sums
are convergent and have no regularization ambiguity. The only
ambiguous terms, defined modulo an additive constant, are the ones with 
$a=b$. We will comment on the impact of such a regularization
ambiguity in the sequel.

The potential $V_{ab} (x)$ is a modular function on the complex torus 
$(\ba,\bb)$ with quasiperiodicity
\begin{eqnarray}
V_{ab} (x+\ba) &=& e^{i\phi_{ab}} \, V_{ab} (x) \nonumber \\
V_{ab} (x+\bb) &=& e^{i\theta_{ab}} \, V_{ab} (x) 
\end{eqnarray}
It has a double pole at $x=0$, with principal part
\be
V_{ab} (x) = \frac{1}{x^2} + O( x^0 )
\ee
and no other poles in each cell. These properties uniquely define
$V_{ab}$ and allow for an expression in terms of theta-functions.
We put
\be
V_{ab} (x) = A e^{i\frac{x}{\ba}\phi_{ab}} \frac{
\tha \left( \frac{\pi}{\ba} (x-q_1) \right)
\tha \left( \frac{\pi}{\ba} (x-q_2)\right)}
{\tha \left( \frac{\pi}{\ba} x \right)^2}
\label{VAqq}
\ee
where $q_{1,2}$ are the as yet unknown zeros of $V_{ab} (x)$
and the theta-functions appearing above have complex period
$T = \bb/\ba$. This has the right quasiperiodicity under
$x \to x+\ba$. In order to also have the right quasiperiodicity under
$x \to x+\bb$, $q_{1,2}$ must satisfy
\be
q_1 + q_2 = \frac{1}{2\pi} (\ba \theta_{ab} - \bb \phi_{ab} )
\equiv Q_{ab}
\label{qqQ}
\ee
and to have the right behavior around $x=0$ we must further have
\be
A = \frac{\pi^2 \tha' (0)^2}{\ba^2 
\tha \left( \frac{\pi q_1}{\ba} \right)
\tha \left( \frac{\pi q_2}{\ba} \right)} 
\label{Aqq} 
\ee
\be
\frac{ \tha' \left( \frac{\pi}{\ba} q_1 \right) }
{ \tha \left( \frac{\pi}{\ba} q_1 \right) } +
\frac{ \tha' \left( \frac{\pi}{\ba} q_2 \right) }
{ \tha \left( \frac{\pi}{\ba} q_2 \right) }
= i\frac{\phi_{ab}}{\pi}
\label{qqp}
\ee
The equations (\ref{qqQ}) and (\ref{qqp}) above determine
$q_1$ and $q_2$, while (\ref{Aqq}) then determines $A$. It may be
possible to express $q_1$, $q_2$ in a more explicit form, or to
recast (\ref{VAqq}) in a form more symmetric in $\ba$,$\bb$, by 
using theta-function identities. Finally, the self-coupling
${\tilde V}_{ab}$ can be extracted from $V_{ab} (x)$ as
\be
{\tilde V}_{ab} = \lim_{x \to 0} \left( V_{ab} (x) - \frac{1}{x^2}
\right)
\label{Vtil}
\ee

To sum up, we obtain a $U(n)$ non-invariant spin-generalization of the
elliptic model given by a hamiltonian of the form (\ref{Hab}) but with
the potentials appearing now being given by (\ref{VAqq},\ref{Vtil}).
The $U(n)$ invariance of the original model is, again, broken
down to the diagonal abelian sungroup $U(1)^n$ due to the dependence of
the potential on $a,b$.
The $U(n)$-invariant spin-Weierstrass model is regained for $\phi_{ab}
= \theta_{ab} =0$, that is, trivial matrices $U$ and $V$. 

We point out that for $\theta_{ab} = \phi_{ab} =0$, that is,
$Q_{ab} =0$, the equations for $q_{1,2}$ (\ref{qqQ},\ref{qqp})
are satisfied for {\it any} $q_1 = -q_2$ leading to an apparent
arbitrariness. As can be seen, however, by applying the addition 
formula
\be
\tha (x+q) \tha (x-q) \thh (0)^2 =  \tha (x)^2 \thh (q)^2
- \thh (x)^2 \tha (q)^2
\ee
this simply amounts to an arbitrary additive constant to the
expression for $V_{ab} (x) \equiv V (x)$. This corresponds to the
need for regularization for this expression in the absence of
phases, as explained before.
(In the case of the Weierstrass function this is fixed by further
requiring that the $O( x^0 )$ part of the function at $x=0$ vanish,
which picks $q=\pi T/2$ and makes $\thh (q)$ above vanish.)
We also point out that we can pick {\it any} of these values
for $q_1 = -q_2$ at the limit $Q=0$ by appropriately choosing
the ratio $\phi_{ab} / \theta_{ab}$ as they both go to zero.

The ambiguity of the terms with $a=b$ can be fixed in the same
way: we can choose phases $\phi_{aa} \neq 0$, $\theta_{aa} \neq 0$,
evaluate the expressions, and then let $\phi_{aa} , \theta_{aa} \to 0$.
This will lead to arbitrary additive constants $C_a$, depending
on the ratio $\phi_{aa} / \theta_{aa}$ as we take them to zero.
The same constants, however, will appear in both $V_{aa} (x)$
and ${\tilde V}_{aa}$. Their net contribution to the hamiltonian
will be
\be
\Delta H = 
\half \sum_{i \neq j} \sum_a C_a \, (S_i)_{aa} (S_j)_{aa} +
\half \sum_i \sum_a C_a \, (S_i)_{aa} (S_i)_{aa} =
\half \sum_a C_a \, (S_{aa})^2
\ee
Since the diagonal components of the total spin $S$ are still
constants of the motion, due to the residual $U(1)^n$ invariance,
this amounts to the addition of an overall constant, and thus leads
to systems that are trivially related.

The same discussion applies if the angles $\phi_a$ and $\theta_a$
coincide for two or more values of $a$ belonging to a subspace
of indices $I$, in which case $\phi_{ab} = \theta_{ab} =0$ for 
$a,b \in I$. This will result to a
constant additive matrix $C_{ab}$ in the potential for this subspace
of indices $I$, leading to an extra contribution to the hamiltonian
\be
\Delta H = \sum_{a,b \in I} C_{ab} \, S_{ab} S_{ba}
\ee
Since $\phi_{ab} = \theta_{ab} =0$ for $a,b \in I$, however, the 
corresponding
subgroup of $U(n)$ remains unbroken, and thus the corresponding
components of the total spin $S_{ab}$ appearing above are constants
of the motion. Once again, the arbitrary terms are constant and we
essentially obtain a unique system.

Overall, this is a generalization of the spin-Weierstrass model
to one involving $2n$ phases that break the $U(n)$ invariance
and promote the potential to a modular function. The potential lives
on a complex torus in the coordinates, where translations around each 
nontrivial
cycle are accompanied by spin transformations. The model obtained
in (f) can be though of as the limit of the present model with
$\bb \to \infty$, in which case $\theta_a$ become
irrelevant. The properties of this modular potential and of the
corresponding new integrable model deserve further study.

\vskip 0.2cm

We conclude by commenting on the quantum mechanical extensions
of the new models introduced here. All known classical models of 
Calogero type are also integrable quantum mechanically. We expect,
therefore, that the models introduced here will also have integrable
quantum mechanical counterparts. This is indeed the case at least
with the model (f), which was fully solved quantum mechanically
in \cite{AP}. The quantum version of these models, however,
does not seem to be
directly accessible by the method used here. The reason is that, in
general, the constraints implied by the restrictions are second class.
Therefore, we cannot carry over the solution of the quantum model
in the unrestricted space and apply the constraints as operator
relations on the Hilbert space. The proper quantum mechanical
treatment of these new models remains, therefore, an interesting
open issue.

\end{document}